\documentclass[a4paper,11pt]{article}
\pdfoutput=1 % if your are submitting a pdflatex (i.e. if you have
\usepackage{jcappub} % for details on the use of the package, please
\usepackage[T1]{fontenc} % if needed
\usepackage{caption}

\newcommand{\beq}{\begin{equation}}
\newcommand{\eeq}{\end{equation}}
\newcommand{\bea}{\begin{eqnarray}}
\newcommand{\eea}{\end{eqnarray}}
\newcommand{\beqn}{\begin{equation*}}
\newcommand{\eeqn}{\end{equation*}}
\newcommand{\bean}{\begin{eqnarray*}}
\newcommand{\eean}{\end{eqnarray*}}

\newcommand*{\cref}[1]{Chapter~\ref{#1}}

\title{\boldmath Quartic hilltop inflation revisited}

\author[a]{Gabriel Germ\'an}
\affiliation[a]{Instituto de Ciencias F\'{\i}sicas, Universidad Nacional Aut\'onoma de M\'exico, Av. Universidad s/n, Cuernavaca, Morelos 62210, Mexico}
\emailAdd{gabriel@icf.unam.mx}
\abstract{We implement a procedure by which the parameters present in the potential of Quartic Hilltop Inflation (QHI) are eliminated in favor of the scalar spectral index $n_s$ and the tensor-to-scalar ratio $r$. By doing this it is posible to obtain in a straightforward and simple way the equations of a previous analysis where an analytical treatment of QHI in the large field limit is given. 
This procedure also allows a more precise discussion of general properties of the model. Also, using a constraint from the reheating epoch it is possible to find bounds for the parameters of the model as well as for quantities of interest such as the running of the scalar index, the reheating temperature and the inflationary scale. Since the bounds found come from expressions given exclusively in terms of $n_s$ and $r$ they will continue to narrow as the measurements of the observables $n_s$ and $r$ become more sensitive.}

\begin{document}
\maketitle
\flushbottom

%%%%%%%%%%%%%%%%%%%%%%%%%%%%%%%%
\section {\bf Introduction}\label{INT}

Quartic Hilltop Inflation (QHI) is an inflationary model (for reviews on inflation see e.g., \cite{Linde:1984ir}-\cite{Martin:2018ycu}) that has received renewed interest recently. Originally this model was studied as a small field model \cite{Linde:1981mu}, \cite{Kinney:1995cc}, \cite{Boubekeur:2005zm} with results incompatible with observations. Subsequently, this model, in a large field regime  \cite{Martin:2013tda}-\cite{Lin:2019fdk}, has been compatible with the observations to such a degree that it is considered both in the {\it Encyclop{\ae}dia Inflationaris}  \cite{Martin:2013tda} and in the {\it Planck 2018} article  \cite{Akrami:2018odb} as a viable model appearing prominently in one of the main figures of the last paper. Although QHI is not without criticism \cite{Kallosh:2019jnl} it is interesting enough to warrant further study.
In a recent and interesting article \cite{Dimopoulos:2020kol}, QHI has been studied in the large field limit from an analytical perspective in an effort to complement and clarify numerical work previously developed by other authors \cite{Martin:2013tda}-\cite{Kallosh:2019jnl}. The analytical work carried out in \cite{Dimopoulos:2020kol} culminates with the obtention of a formula that relates the tensor-to-scalar ratio $r$ with the scalar spectral index $n_s$, allowing an exact reproduction of the numerical curves for  $N_{ke}=50$ and  $N_{ke}=60$ presented in the Planck 2018 article  \cite{Akrami:2018odb}, where $N_{ke} \equiv -\frac{1}{M^2}\int_{\phi_k}^{\phi_e}\frac{V}{V'}d\phi$ is the number of e-folds of expansion  of the universe from $\phi_k$ up to the end of inflation at $\phi_e$. 

The purpose of this work is twofold: on the one hand, it aims to show in this concrete example how the procedure consisting on eliminating the parameters of the model (appearing in the potential) in terms of the observables $ n_s $ and $ r $ can quickly lead us to obtain important results in a clear and direct way. On the other hand, the same procedure when applied using constraints from the reheating era (for reviews on reheating see e.g., \cite{Bassett:2005xm},  \cite{Allahverdi:2010xz}, \cite{Amin:2014eta}) allows us to find bounds for quantities of interest written in terms of $ n_s $ and $ r $ directly by the application of the bounds for these observables. The bounds for both $n_s$ and $r$ will keep tightening as more sensitive measurements are carried out, and so will all quantities written in terms of them.

For future reference we give here the expressions for spectral indices and observables at first order in the slow-roll (SR) approximation (see e.g.  \cite{Lyth:1998xn}, \cite{Liddle:1994dx})
\begin{eqnarray}
n_{t} &=&-2\epsilon = -\frac{r}{8} , \label{Int} \\
n_{s} &=&1+2\eta -6\epsilon ,  \label{Ins} \\
%n_{tk} &=&4\epsilon\left( \eta -2\epsilon\right), \label{Intk} \\
n_{sk} &=&16\epsilon \eta -24\epsilon ^{2}-2\xi_2, \label{Insk} \\
A_s(k) &=&\frac{1}{24\pi ^{2}} \frac{V}{%
M^4\epsilon},
\label{IA} 
\end{eqnarray}
where $n_{s}$ is the the scalar index and $n_{sk} \equiv \frac{d n_{s}}{d \ln k}$ its running, usually denoted by $\alpha$. Here we prefer to use this more symmetrical notation between scalar and tensorial quantities. The amplitude of scalar density perturbations at wave number $k$ is $A_s(k)=2.1\times 10^{-9}$. All quantities in Eqs.~\eqref{Int} to  \eqref{IA} are evaluated at the horizon crossing scale of wavenumber mode $k$.
The SR parameters appearing above are defined by
\begin{equation}
\epsilon \equiv \frac{M^{2}}{2}\left( \frac{V^{\prime }}{V }\right) ^{2},\quad\quad
\eta \equiv M^{2}\frac{V^{\prime \prime }}{V}, \quad\quad
\xi_2 \equiv M^{4}\frac{V^{\prime }V^{\prime \prime \prime }}{V^{2}}.
\label{Spa}
\end{equation}
Also,  $M$ is the reduced Planck mass $M=2.44\times 10^{18} \,\mathrm{GeV}$ and primes on $V$ denote derivatives with respect to the inflaton field $\phi$. 

The organization of the article is as follows: In Section \ref{QHIR} we reproduce in an almost trivial way the two main equations of \cite{Dimopoulos:2020kol}. This is done by eliminating the parameters of the model in terms of the observables $n_s$ and $r$. We also show how the presence in $N_{ke}$ of terms coming from the end of inflation is not only necessary but inevitable and obtain a first order formula relating the scalar index $n_s$ with the number of e-folds during inflation.
In Section \ref{CONS} we use the constraint $N_{re}(\omega_{re})\geq 0$ from reheating to find constraints for $n_{s}$ and/or $r$ and in turn use these to constrain quantities written in terms of them. In this way we find bounds for the parameters of the model, the reheating temperature at the end of inflation, the running index and the scale of inflation. The quantity $N_{re}(\omega_{re})$ above is the number of e-folds during reheating and $\omega_{re}$ is the equation of state parameter (EoS). At the end of the section we collect in Table~\ref{bounds} the main results of this section. 
 Finally we conclude in Section \ref{CON}.

%%%%%%%%%%%%%%%%%%%%%%%%%%%%%%%%
%%%%%%%%%%%%%%%%%%%%%%%%%%%%%%%%
%%%%%%%%%%%%%%%%%%%%%%%%%%%%%%%%
%%%%%%%%%%%%%%%%%%%%%%%%%%%%%%%%
\section {\bf Quartic hilltop inflation revisited }\label{QHIR} 

In what follows we are interested in the large-field (small $\lambda$) limit of QHI whose potential is given by
\begin{equation}
V=V_0\left(1-\lambda\left(\frac{\phi}{M}\right)^4+\cdot\cdot\cdot \right)\;.
\label{pot}
\end{equation}
An expression for $\phi_k$, the inflaton at horizon crossing, is obtained  by solving Eq.~\eqref{Ins} 
\begin{equation}  
\delta_{n_s}+2\eta-\frac{3}{8}r=0\;,
\label{spectral}
\end{equation}
where $\delta_{n_s}$ is defined as $\delta_{n_s}\equiv 1-n_s$ and $\eta=-\frac{12\lambda \phi^2}{M^2\left(1-\lambda \left(\frac{\phi}{M}\right)^4\right)}$, with the result at $\phi=\phi_k$
\begin{equation}
\left(\frac{\phi_k}{M}\right)^2=\left(\frac{96}{8\delta_{n_s}-3r}\right)\left(-1+\sqrt{1+\frac{1}{\lambda}\left(\frac{8\delta_{n_s}-3r}{96}\right)^2}\right), \quad r<\frac{8}{3}\delta_{n_s}\;.
\label{ficrossing}
\end{equation}
We can eliminate the parameter $\lambda $ substituting $\phi_k$ in Eq.~\eqref{Int}, $r=16\epsilon$, and solving for $\lambda$ 
\begin{equation}
\lambda=\frac{(8\delta_{n_s}-3r)^4}{27648r(16\delta_{n_s}-3r)}\;.
\label{lambada}
\end{equation}
This is  Eq.(31) of \cite{Dimopoulos:2020kol} but written in terms of the two observables $n_s$ and $r$ instead of $n_s$ and $N_{ke}$, where $N_{ke}$ is the number of e-folds from $\phi_k$ up to the end of inflation at $\phi_e$. To obtain  $\lambda=\lambda(n_s,N_{ke})$ we substitute  $r$ in Eq.~\eqref{lambada}  with $r$ given  by Eq.~\eqref{rs} below.

\begin{figure}[tb]
\captionsetup{justification=raggedright,singlelinecheck=false}
\par
%\begin{center}
\includegraphics[width=12cm]{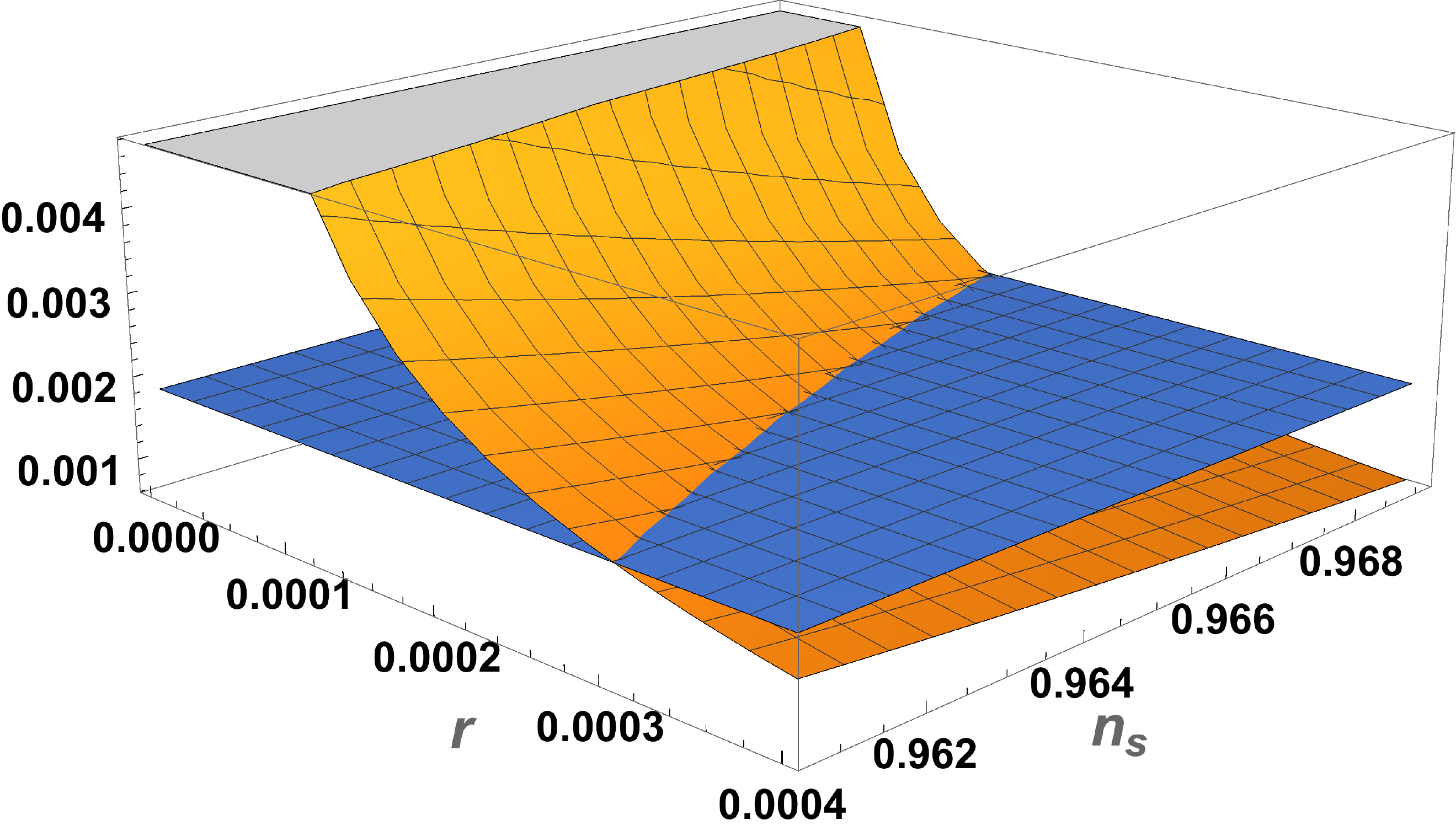}
\caption{\small Plot of the parameter $\lambda$ as a function of $n_s$ and $r$. The horizontal plane corresponds to the limiting value given by $\lambda=1/540\approx 1.85\times 10^{-3}$ which separates the two conditions for ending inflation: for $\lambda <1.85\times 10^{-3}$ we should find $\phi_e$ by solving $\epsilon=1$ (see Eq.~\eqref{fieeeps}) while the case $\lambda >1.85\times 10^{-3}$ requires solving the condition $\eta=-1$.
}
\label{lambda}
%\end{center}
\end{figure}

The end of inflation is given by  the condition $\epsilon=1$ in one case and also by the condition $\eta=-1$ in a second possible case. To find the value of $\lambda$ at which both conditions coincide and thus separate them we solve the equation $\epsilon=-\eta$ with the result $\phi=(3/5\lambda)^{1/4}$. With this value for $\phi$ we substitute it in any of the conditions ($\epsilon=1$ or $\eta=-1$) and solve for $\lambda$ obtaining its limiting value $\lambda_{l}$ (see Fig.\,\ref{lambda})
\begin{equation}
\lambda_{l}=\frac{1}{540}\approx 1.85\times 10^{-3}\;.
\label{lambdalim}
\end{equation}
Thus for $\lambda <1.85\times 10^{-3}$ we should find $\phi_e$ by solving $\epsilon=1$ while the case $\lambda >1.85\times 10^{-3}$ requires solving the condition $\eta=-1$.
The solution to $\epsilon=1$ is given by
\begin{equation}
\frac{\phi_e}{M}=\frac{1}{(6\lambda)^{1/4}} \left(2\sqrt{6\lambda}+\sqrt{2+24\lambda+c_1}-\sqrt{4+48\lambda-c_1+\frac{12\sqrt{6\lambda}(1+8\lambda)} {\sqrt{2+24\lambda+c_1}}}\right)^{1/2}\;,
\label{fieeeps}
\end{equation}
where $c_1=\frac{2+2^{1/3}\left(2+27\lambda+3\sqrt{3\lambda(4+27\lambda)}\right)^{2/3}}{\left(1+27\lambda/2+3/2\sqrt{3\lambda(4+27\lambda)}\right)^{1/3}}$. For small $\lambda$ we have
\begin{equation}
\frac{\phi_e}{M}=\frac{1}{\lambda^{1/4}} -\frac{1}{\sqrt{2}}+\frac{3}{4}\lambda^{1/4}+\cdot \cdot \cdot\;.
\label{fieeepsapp}
\end{equation}
%%%%%%%%%%%%%
%%%%%%%%%%%%%
The solution to $\eta=-1$ is given by $\phi_{e\eta}/M=(-6+\sqrt{(1+36\lambda)/\lambda}\,)^{1/2}$.

The end of inflation at leading order for small $\lambda$ is $\phi_e/M\approx \lambda^{-1/4}$. Thus, the number of e-folds during inflation, $N_{ke} = -\frac{1}{M^2}\int_{\phi_k}^{\phi_e}\frac{V}{V'}d\phi$, can be calculated with the result
\begin{equation}
N_{ke}=\frac{24\left(8\delta_{n_s}-\sqrt{3r(16\delta_{n_s}-3r)}\right)}{(8\delta_{n_s}-3r)^2}\;.
\label{Nkefolds}
\end{equation}
Solving Eq.~\eqref{Nkefolds} for $r$ in terms of $\delta_{n_s}$ and $N_{ke}$ we get the solution
\begin{equation}
r=\frac{8}{3}\delta_{n_s}\left(1-\frac{\sqrt{6N_{ke}\delta_{n_s}-9}}{N_{ke}\delta_{n_s}} \right)\;,
\label{rs}
\end{equation}
which clearly satisfies the condition $r<\frac{8}{3}\delta_{n_s}$ coming from Eq.~\eqref{ficrossing}. This solution is exactly Eq.~(35) of \cite{Dimopoulos:2020kol} (see also \cite{Lin:2019fdk}). The reader will appreciate the effortless way our approach has reproduced the two main equations of Ref.~\cite{Dimopoulos:2020kol}. The trick consists in eliminating the parameters of the model in terms of the observables. In this case, $\lambda$ and  $V_0$ in terms on $n_s$ and $r$ although $V_0$  does not participate in the determination of $r$. Because both $n_s$ and $r$ are observables their bounds will keep tightening as more sensitive measurements are carried out, and so will all quantities written in terms of them.
%%%%%%%%%%%%%%
\begin{figure}[t!]
\captionsetup{justification=raggedright,singlelinecheck=false}
\par
\begin{center}
\includegraphics[trim = 0mm  0mm 1mm 1mm, clip, width=7.5cm, height=5.cm]{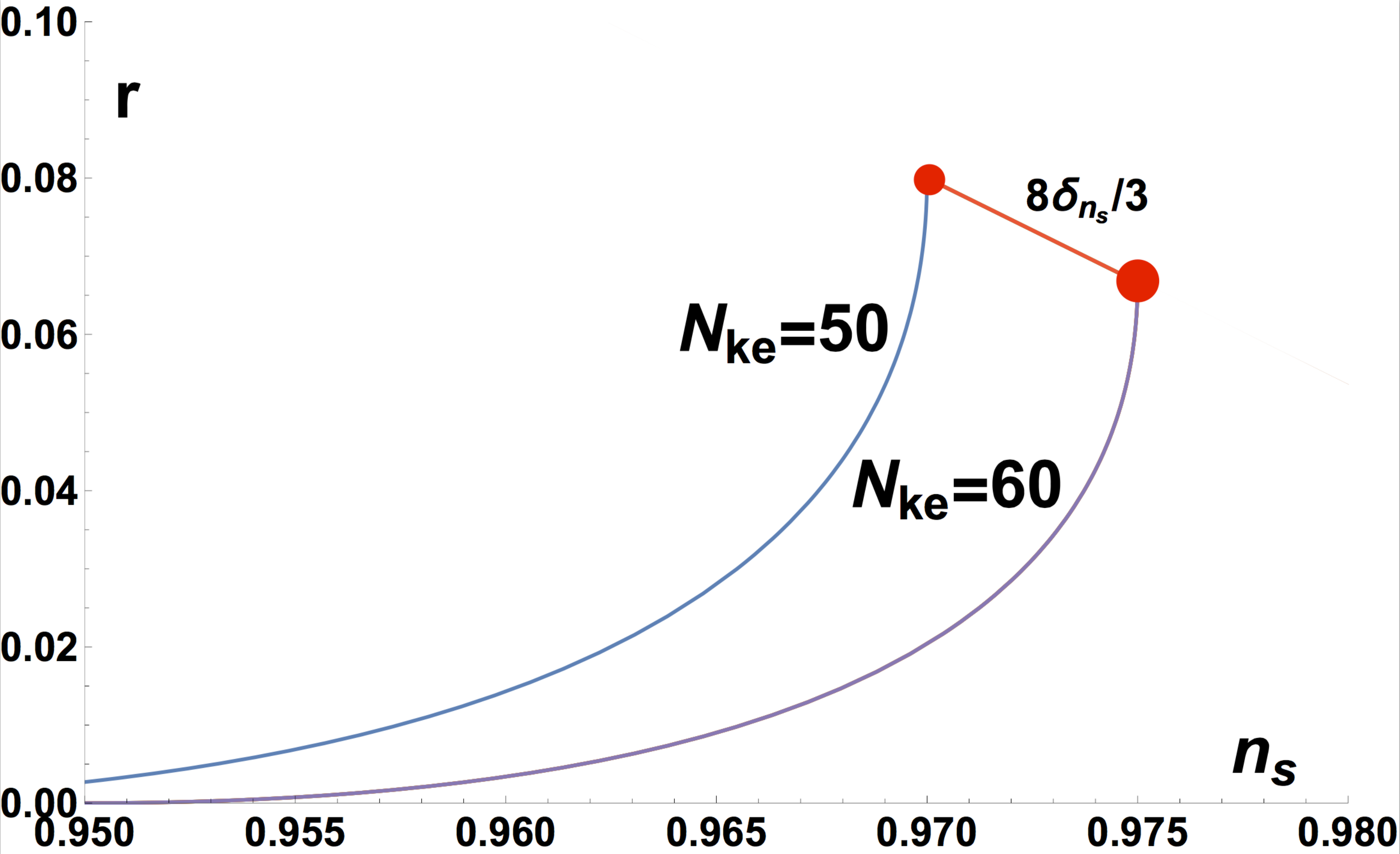}
\includegraphics[trim = 0mm  0mm 1mm 1mm, clip, width=7.5cm, height=5.cm]{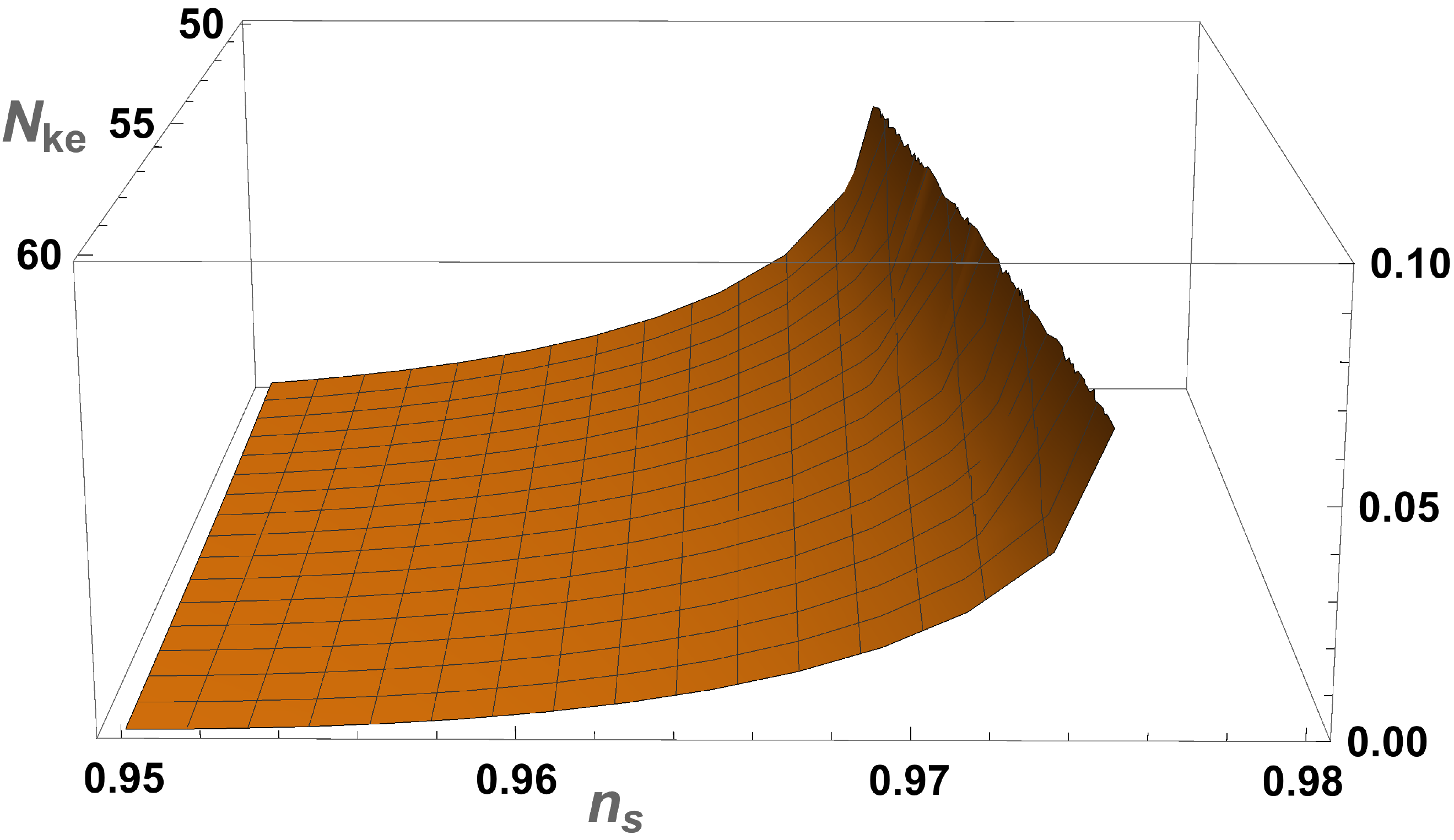}
\end{center}
\caption{The l.h.s. panel plots $r$ from Eq.~\eqref{rs} as a function of $n_s$ for the cases $N_{ke}=50$ and $N_{ke}=60$ e-folds of expansion during the inflationary epoch ending in linear inflation where $r=8 \delta_{n_s}/3$ and $\delta_{n_s}$ is defined by $\delta_{n_s}\equiv 1-n_s$. As discussed in the text this is an unphysical limit, in agreement with \cite{Kallosh:2019jnl}. The r.h.s. is a trivial extension of the previous figure in the $N_{ke}$ direction. The l.h.s figure should be compared with Fig.~8 of the Planck Collaboration 2018 article \cite{Akrami:2018odb} where quartic inflation is considered (reproduced in Fig.~\ref{Pl} below).
}
\label{r}
\end{figure}
%%%%%%%%%%%%%
We show in Fig.\,\ref{r} the solution given by Eq.~\eqref{rs} for  $N_{ke}=50$ and $N_{ke}=60$ and also its trivial extension in the $N_{ke}$ direction. We see that the solution ends in linear inflation where $r=\frac{8}{3}\delta_{n_s}$, this occurs for $\delta_{n_s}$ reaching the value $\frac{3}{2}\frac{1}{N_{ke}}$, where the radicand in Eq.~\eqref{rs} vanishes. 
We also see that in the same limit when $r=\frac{8}{3}\delta_{n_s}$ both $\phi_k$ and $\phi_e$ given by Eqs.~\eqref{ficrossing} and \eqref{fieeeps} respectively, diverge. To appreciate this we eliminate from Eq.~\eqref{ficrossing} the parameter $\lambda$ obtaining $\phi_k/M=(12\sqrt{2r})/(8\delta_{n_s}-3r)$.
Thus, the linear regime limit is not physical as discussed in \cite{Kallosh:2019jnl}. Fig.\,\ref{r} should be compared with the Fig.~8 given by the Planck 2018 article  \cite{Akrami:2018odb} and reproduced here in Fig.\,\ref{Pl}. 
%%%%%%%%%%%%%
\begin{figure}[tb]
\captionsetup{justification=raggedright,singlelinecheck=false}
\par
%\begin{center}
\includegraphics[width=12cm]{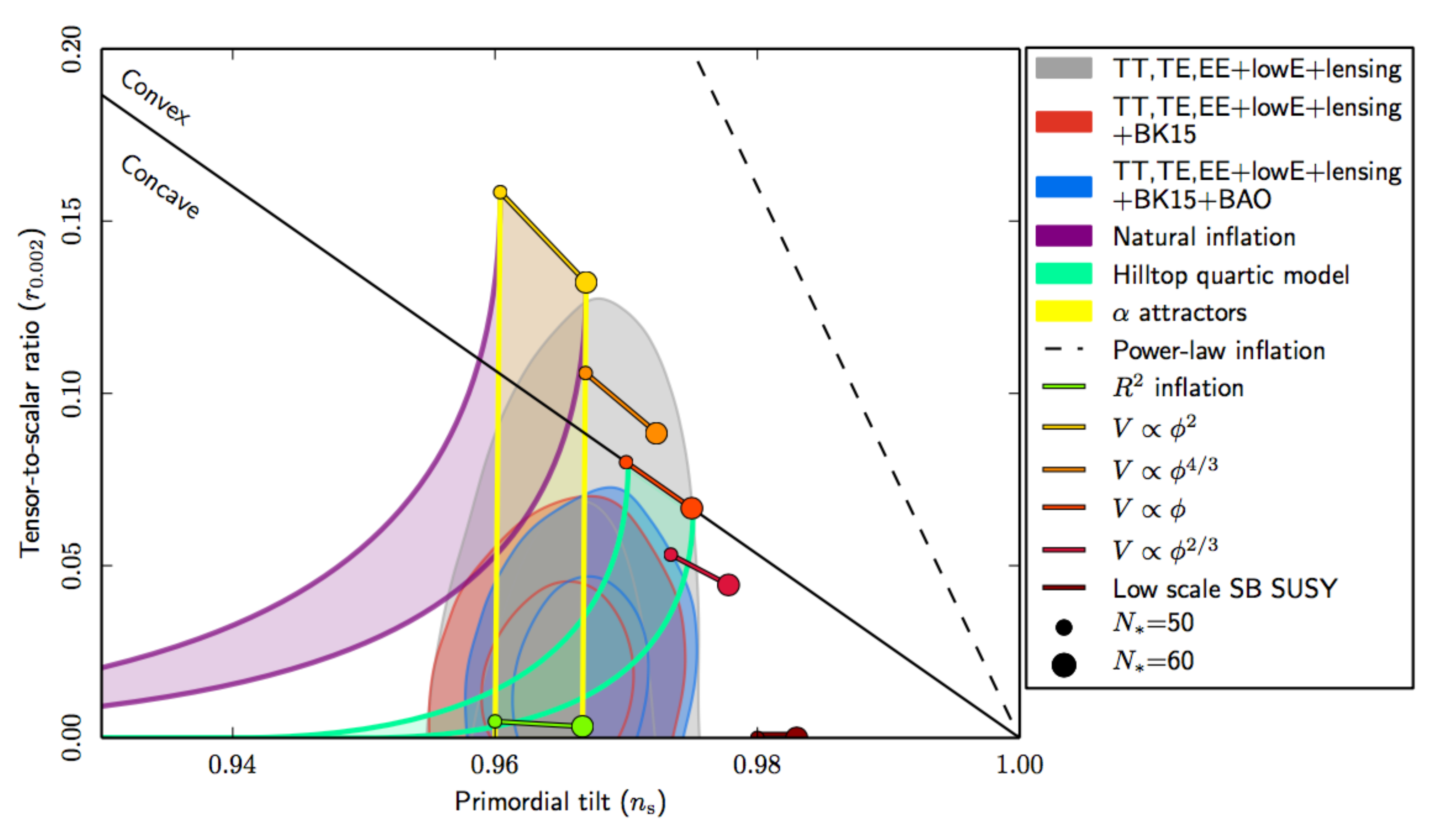}
\caption{\small This is Fig.~8 of the Planck Collaboration 2018 article \cite{Akrami:2018odb} where quartic inflation is considered together with several other interesting models of inflation (see description in the right hand side panel of the figure). From the figure we see that there is a substantial overlap of the predictions of quartic inflation with Planck alone and in combination with BK15 or BK15+BAO data.}
\label{Pl}
%\end{center}
\end{figure}
%%%%%%%%%%%%%

To obtain the leading dependence of $n_s$ on the number of e-folds during inflation we rewrite Eq.~\eqref{rs} as an equation for $n_s$, then we find
\begin{equation}
n_s=1-\frac{3}{8}\left(4 N_{ke}+r-2\sqrt{2N_{ke}(2N_{ke}+r)-8} \right)\approx 1-\frac{3}{2}\frac{1}{N_{ke}}\;,
\label{nsrs}
\end{equation}
Earlier estimates \cite{Lyth:1998xn} in the small field (large $\lambda$) limit give the result $n_s\approx 1-\frac{3}{N_{ke}}$. Thus, taking e.g., $N_{ke}=50$ the large-field (small $\lambda$) limit studied here gives $n_s\approx 0.97$ while the old estimate based on the small field (large $\lambda$) limit gives $n_s \approx 0.94$. Note that the requirement of a positive radicand in Eq.~\eqref{rs} take us immediately to the result \eqref{nsrs} above.

\begin{figure}[tb]
\captionsetup{justification=raggedright,singlelinecheck=false}
\par
%\begin{center}
\includegraphics[width=12cm]{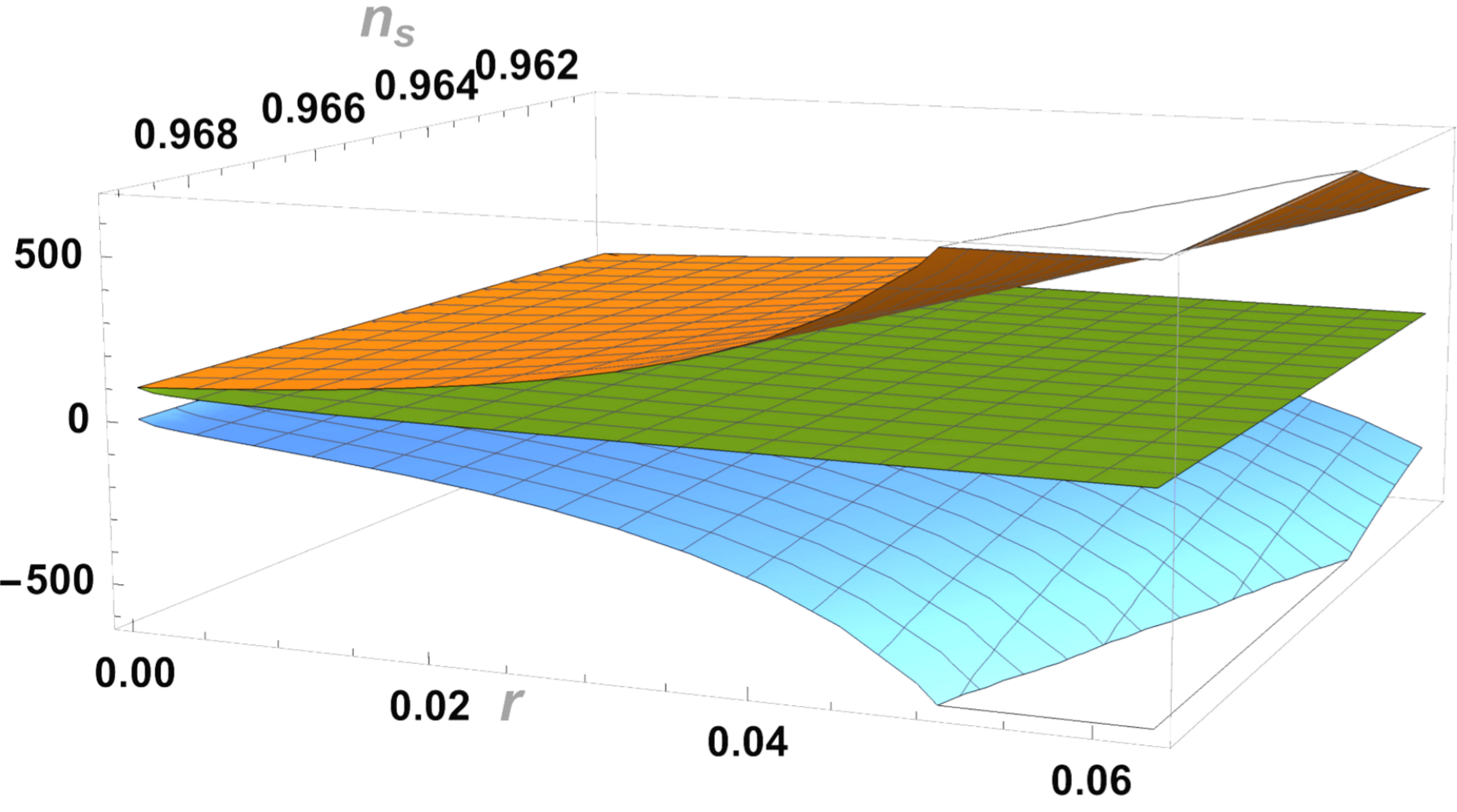}
\caption{\small Plot of the number of e-folds during inflation $N_{ke} = -\int_{\phi_k}^{\phi_e}\frac{V}{V'}d\phi $ written as $N_{ke}=N_{k}+N_{e}$ where $N_{k}\equiv \frac{\phi_k^2}{8M^2}+\frac{\phi_k^{-2}}{8\lambda M^{-2}}$ (top brown sheet) and $N_{e}\equiv -\frac{\phi_e^2}{8 M^2}-\frac{\phi_e^{-2}}{8\lambda M^{-2}}$ (bottom blue sheet). $N_{ke}$ is given by the green sheet (in between). The number of e-folds during inflation $N_{ke}$ can also be written as $N_{ke} =\frac{1}{8 M^2}(\phi_k^2-\phi_e^2)+ \frac{1}{8\lambda M^{-2}}(\phi_k^{-2}-\phi_e^{-2})$ thus, neglecting $\phi_e^{-2}$ in favor of $\phi_k^{-2}$ in the second term is equivalent to neglecting $\phi_k^{2}$ in favor of $\phi_e^{2}$ in the first. The contribution from the end of inflation $N_{e}$ to the the total number of e-folds during inflation $N_{ke}$ is not only important but cannot be avoided.}
\label{enes1}
%\end{center}
\end{figure}

The importance of the end of inflation for the result \eqref{rs} can be understood as follows:
the expression for the number of e-folds during inflation given by $N_{ke} = -\frac{1}{M^2}\int_{\phi_k}^{\phi_e}\frac{V}{V'}d\phi$  is formed by adding two terms together i.e., $N_{ke}=N_{k}+N_{e}$ where $N_{k}\equiv \frac{\phi_k^2}{8 M^2}+\frac{\phi_k^{-2}}{8\lambda M^{-2}}$ and $N_{e}\equiv -\frac{\phi_e^2}{8M^2}-\frac{\phi_e^{-2}}{8\lambda M^{-2}}$. These quantities are plotted in Fig.\,\ref{enes1} where the lower sheet (blue) corresponds to $N_{e}$, the upper sheet (brown) to $N_{k}$ and the sum $N_{ke}$ is given by the sheet in between (green). From the figure we see that $N_{k}$ is not a good approximation to the number of e-folds during inflation even for very small $r$. Even more to the point, writing the expression for $N_{ke}$ as $N_{ke} =\frac{1}{8M^2}(\phi_k^2-\phi_e^2)+\frac{1}{8\lambda M^{-2}}(\phi_k^{-2}-\phi_e^{-2})$ it is immediately obvious that neglecting $\phi_e^{-2}$ in favor of $\phi_k^{-2}$ in the second term is equivalent to neglecting $\phi_k^{2}$ in favor of $\phi_e^{2}$ in the first. Thus, the contribution from the end of inflation $N_{e}$ to the total number of e-folds during inflation $N_{ke}$ is not only important but cannot be avoided. Note that no particular value of $\phi_e$ is required for the previous argument.
%%%%%%%%%%%%%
%%%%%%%%%%%%%
%%%%%%%%%%%%%%%%%%%%%%%%%%%%%%%%
%%%%%%%%%%%%%%%%%%%%%%%%%%%%%%%%
\section {\bf Constraints from reheating}\label{CONS}
Having shown the economics of our procedure we now  impose bounds to several quantities of interest during inflation coming from constraints during the reheating era. 
To decide on possible values for $\lambda$ we now study constraints from reheating. As we can see from Eq.~\eqref{pot}, however, the potential does not contain a region where reheating can be described. One usually expects to see a convex potential with a minimum around which oscillations of the inflaton take place and reheating occurs. We will assume that it is possible to find some extra terms responsible for creating a minimum in such a way that the inflationary sector is well approximated by the existing terms in Eq.~\eqref{pot}. The missing part of the potential where we assume that reheating occurs is in a way parameterized by $\omega_{re}$. For potentials $V\sim \phi^n$ around the minimum where oscillations take place, an inflaton oscillating with frequency~$\sim a^{3(n-2)(n+2)}$ has an EoS given by \cite{Turner:1983he}
\begin{equation}
\omega = \frac{n-2}{n+2} \;.
\label{EoS}
\end{equation}
Here, however, in the absence of a minimum for the potential we should look for results independent of the particular value of $\omega_{re}$. Thus, we will be happy to consider the very general constraint of having a greater than or equal to zero number of e-folds during reheating, $N_{re}(\omega_{re})\geq0$. The particular value $N_{re}= 0$ corresponds to instantaneous reheating and is independent of the value of $\omega_{re}$.

We assume that the reheating epoch is on average characterized by an EoS, denoted by $\omega_{re}$, which can take values larger than $-1/3$. Building on previous work \cite{Liddle:2003as}, \cite{Dodelson:2003vq}, \cite{Liddle:1994dx}, even without knowing the details of the potential around its minimum it is possible to find an expression for the number of e-folds during reheating  \cite{Dai:2014jja}, \cite{Munoz:2014eqa}  as follows (see also e.g., section 3 of \cite{German:2020iwg})
\beq
\label{NRE}
N_{re}= \frac{4}{1-3\, \omega_{re}}\left(-N_{k}-\frac{1}{3} \ln[\frac{11 g_{s,re}}{43}]-\frac{1}{4} \ln[\frac{30}{\pi^2 g_{re} } ] -\ln[\frac{\rho^{1/4}_e k}{H_k\, a_0 T_0} ]\right)\;,
\eeq
where  $\rho_{e}$ is the energy density at the end of inflation and $H_k$ is the Hubble function at horizon crossing. The number of degrees of freedom of species during reheating is $g_{re}$ while $g_{s,re}$ is the entropy number of degrees of freedom after reheating. $N_{re}$ can also be written in a more convenient form for our purposes as follows
\beq
\label{NREfinal}
N_{re}=\left(1-3\, \omega_{re}\right)^{-1}\left(\ln\left[\frac{V_k}{V_e}re^{-4N_{k}}
\right]+\ln\left[\frac{\pi^4A_s\, g_{re}}{270} \left(\frac{43}{11\,g_{s,re}}\right)^{4/3} \left(\frac{a_0 T_0}{k}\right)^{4}  \right]\right)\;,
\eeq
where $V_e$ is the potential at the end of inflation and $\rho_e=\frac{3}{2}V_e=\frac{9}{2}\frac{V_e}{V_k}H_k^2M^2=\frac{9\pi^2 A_s M^4}{4}\frac{V_e}{V_k}r$ has been used. Numerical values are given as follows: $g_{s,re}=g_{re}=106.75$, $A_s=2.1\times 10^{-9}$, $T_0=2.725K=9.62\times 10^{-32}$, $k_p=0.05/Mpc=1.31\times 10^{-58}$, the last two quantities are also given by their dimensionless values in Planck units.
All the $n_s$, $r$ dependence is contained in the first term on the r.h.s. of Eq.~\eqref{NREfinal}. To study constraints coming from reheating we can simply study $N_{re}$ numerically by using the solution Eq.~\eqref{fieeeps} to evaluate $V_e$  and $N_k$ however we also give analytical expressions which approximate the problem very well. The overall constant $V_0$ is readily obtained by substituting the potential \eqref{pot} in Eq.~\eqref{IA} and solving for $V_0$ while $V_k$ follows from the same equation after the substitution $\epsilon=r/16$
\begin{equation}
V_{0} = \frac{3A_s\pi^2r(16\delta_{n_s}-3r)}{4(8\delta_{n_s}-3r)}M^4\;.
\label{V0}
\end{equation}
\begin{equation}
V_{k} = \frac{3M^4}{2}A_s \pi^2 r\;.
\label{Vk}
\end{equation}
If one takes only the leading term $\phi_e/M=\frac{1}{\lambda^{1/4}}$ in the expansion of Eq.~\eqref{fieeeps} one gets $V_e=0$ which is clearly of no use for calculating $N_{re}$ so one has to go to the next term in the expansion as in Eq.~\eqref{fieeepsapp} above with the result
\begin{equation}
V_{e} =V_0\left(1-\frac{1}{110592}\left(8\times 3^{3/4}-\frac{8\delta_{n_s}-3r}{r^{1/4}(16\delta_{n_s}-3r)^{1/4}}\right)^4\right)\;.
\label{Ve}
\end{equation}
While $N_{ke}$ is well approximated by using $\phi_e/M = \frac{1}{\lambda^{1/4}}$, $V_e$ requires also of the second term of the small $\lambda$ expansion. In any case, all the numerical results presented here are obtained using the exact solution Eq.~\eqref{fieeeps} and checking that the analytical formulas are indeed good approximations.

A final quantity of physical relevance is the thermalization temperature at the end of reheating
\beq
\label{TRE}
T_{re}=\left( \frac{30\, \rho_e}{\pi^2 g_{re}} \right)^{1/4}\, e^{-\frac{3}{4}(1+\omega_{re})N_{re}}\,,
\eeq
\noindent 
and is given in terms of $N_{re}$ thus, it is convenient to concentrate in what follows on $N_{re}$. 

\begin{figure}[tb]
\captionsetup{justification=raggedright,singlelinecheck=false}
\par
\begin{center}
\includegraphics[width=12cm]{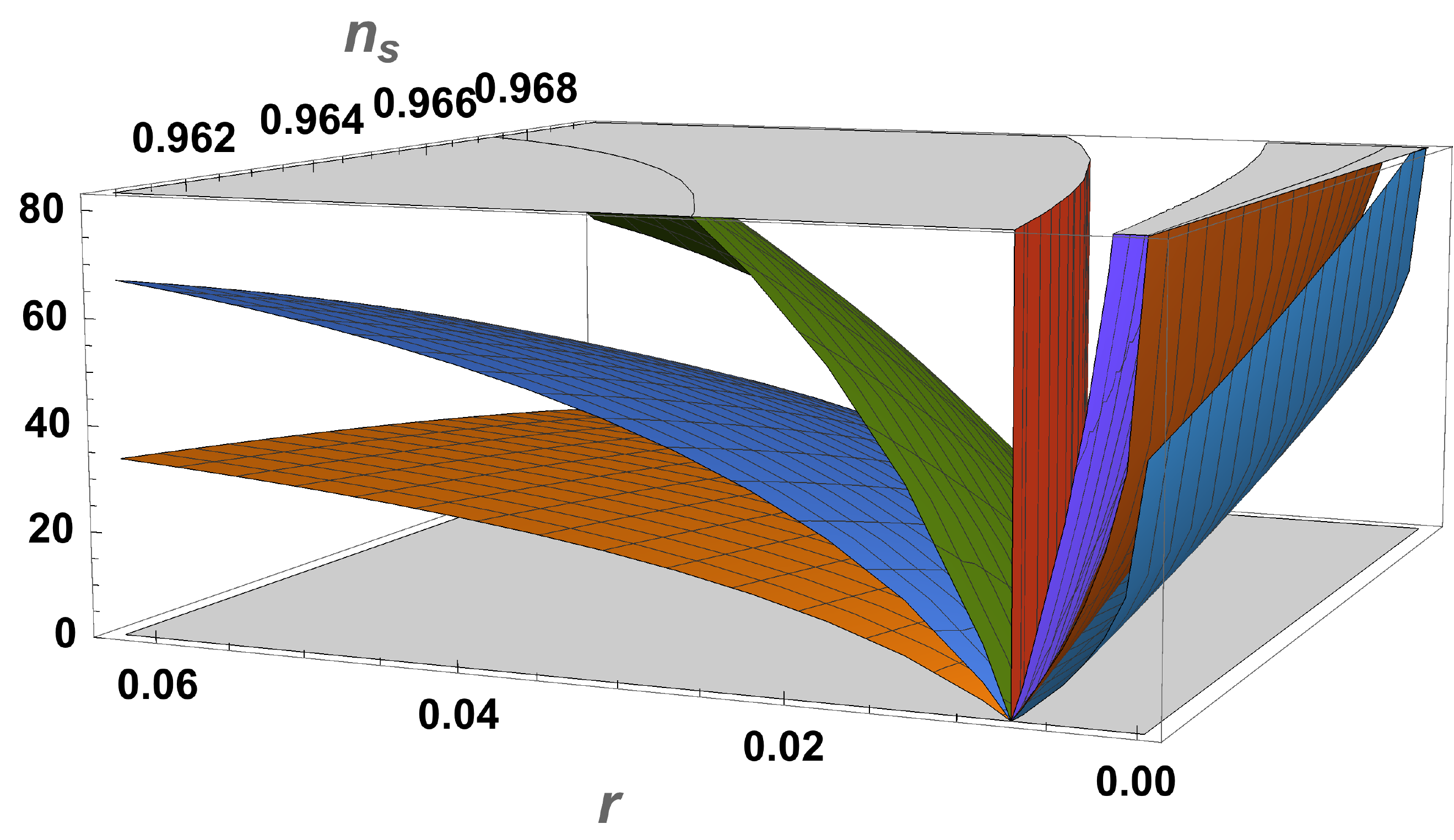}
\caption{\small The number of e-folds during reheating $N_{re}(\omega_{re})$ given by Eq.~\eqref{NREfinal} is shown here as a function of the scalar spectral index $n_s$ and the tensor-to-scalar ratio $r$ for (from left to right) $\omega_{re}=-1/3, 0, 0.17, 0.3332, 0.47, 0.68$ and $1$, respectively. We see that the condition $N_{re}(\omega_{re})\geq 0$ imposes constraints on $r$ with $r>0.0071...$ for $-1/3<\omega_{re}<1/3$ and $r<0.02508...$ when $1/3<\omega_{re}<1$. We can then use these constraints on $r$ to bound the model parameter $\lambda$, the running of the scalar spectral index $n_{sk}$ the thermalization temperature at the end of reheating $T_{re}$ and the scale of inflation $\Delta\equiv V_k^{1/4}$ as given in the Table~\ref{bounds}.}
\label{Nre1}
\end{center}
\end{figure}

In Fig.\,\ref{Nre1} we plot $N_{re}(\omega_{re})$ for (from left to right) $\omega_{re}=-1/3, 0, 0.17, 1/3, 0.47, 0.68, 1$, Eq.~\eqref{NREfinal} is not valid for $\omega_{re}=1/3$ so we actually plot that case for $\omega_{re}=0.3332$. From the general condition $N_{re}\geq 0$ we see that all the EoS in the interval for $-1/3<\omega_{re}<1/3$ (to the left of the vertical sheet) should satisfy that $r>0.00711...$ while the cases with $1/3<\omega_{re}<1$ require $r<0.02508...$. As mentioned above the particular value $N_{re}= 0$ corresponds to instantaneous reheating (see Eq.~\eqref{TRE}) and is independent of the value of $\omega_{re}$, that is why all the sheets in Fig.\,\ref{Nre1} converge in a single line on the $n_s$~vs.~$r$ plane, which is then used to obtain the bounds for $r$ given above. An equivalent way of saying this is that $N_{re}= 0$ occurs because the large round parenthesis in Eq.~\eqref{NREfinal} (which only depends on $n_s$ and $r$) vanishes for any value of $\omega_{re}\neq 1/3$.

Thus, we can actually calculate a bound to the thermalization temperature: in the first case $T_{re}\leq 2.9\times 10^{15} GeV$ while in the second $T_{re}\leq 3.5\times 10^{15} GeV$. In both cases the original range for $n_s$ and the {\it upper} bound for $r$ are as given by the Planck Collaboration 2018 \cite{Aghanim:2018eyx}, \cite{Akrami:2018odb} $0.9607<n_s<0.9691$ and $r<0.063$ but the new bounds for $r$ can now be used to further constrain $\lambda$, thus, for the case $-1/3<\omega_{re}<1/3$ 
\beq
N_{re}(\omega_{re})\geq 0,\quad \Rightarrow\quad  0.0071<r<0.063,\quad \Rightarrow\quad 2.2\times 10^{-8}<\lambda<2.8\times 10^{-5},
\label{caso1}
\eeq
while for $1/3<\omega_{re}<1$
\beq
N_{re}(\omega_{re})\geq 0,\quad \Rightarrow\quad  0<r<0.0251,\quad \Rightarrow\quad 3.0\times 10^{-6}<\lambda.
\label{caso2}
\eeq
We see that in the first case (Eq.~\eqref{caso1}) the end of inflation is dictated by the condition $\epsilon=1$ while in the second both conditions can apply depending on the value of $\lambda$.

%%%%%%%%%%%%%
%%%%%%%%%%%%%
%%%%%%%%%%%%%%%%
 \begin{center}
 \captionsetup{justification=raggedright,singlelinecheck=false}
\par
\addtolength{\tabcolsep}{-7pt}
\begin{table*}[htbp!]
%\begin{ruledtabular}
\begin{tabular}{ccc}
$Characteristic$ & $N_{re}\geq 0$,\quad $-1/3<\omega_{re}<1/3$ & $N_{re}\geq 0$,\quad $1/3<\omega_{re}<1$\\ \hline\\[0.1mm]
$n_s$   &  $0.9607 < n_{s} < 0.9691$  & $0.9607 < n_{s} < 0.9691$\\[2mm] 
$r$   &  $0.0071< r < 0.0630$ & $0< r < 0.0251$\\[2mm]
$\lambda$   &  $2.2\times 10^{-8} < \lambda < 2.8\times 10^{-5} $ & $3.0\times 10^{-6} < \lambda$\\[2mm]
$n_{sk}$  &   $-9.5  \times 10^{-4} <n_{sk} < -3.7\times 10^{-4}$& $-7.3  \times 10^{-4} < n_{sk} < -3.2\times 10^{-4}$\\[2mm]
$T_{re} (GeV)$  & $T_{re}<2.9\times 10^{15}$ & $T_{re}<3.5\times 10^{15}$\\[2mm]
$\Delta (GeV)$   &  $9.4\times 10^{15}< \Delta <1.6\times 10^{16}$ & $0< \Delta <1.3\times 10^{16}$\\[2mm]
\end{tabular}
%\end{ruledtabular}
\caption{\label{bounds} For the EoS $\omega_{re}$ lying in the intervals $-1/3<\omega_{re}<1/3$ and $1/3<\omega_{re}<1$ we show quantities of interest for the quartic hilltop model of inflation. The range for the spectral index $n_s$ comes from the Planck Collaboration 2018,   \cite{Akrami:2018odb}, \cite{Aghanim:2018eyx} data. Starting from the bounds $0<r<0.063$ we impose the condition that the number of e-folds during reheating be greater than or equal to zero $N_{re}(\omega_{re})\geq0$ and from there we find new bounds for the tensor-to-scalar ratio $r$. When $-1/3<\omega_{re}<1/3$, $r$ is bounded from below by 0.0071 and from above (0.0251) when $1/3<\omega_{re}<1$.
Once we have these bounds for $r$ the other quantities like the model parameter $\lambda$, the running of the scalar spectral index $n_{sk}$ the thermalization temperature at the end of reheating $T_{re}$ and the scale of inflation $\Delta\equiv V_k^{1/4}$ are also bounded as shown. As more sensitive measurements are made the bounds for the observables $n_s$ and $r$ will tighten and also the bounds for the other quantities written in terms of them. This is why is important to eliminate the (non-observable) parameters of the model (appearing in the potential) in favor of the observable parameters such as $n_s$ and $r$.}
\end{table*}
\end{center}
%%%%%%%%%%%%%
Using Eqs.~\eqref{Insk} and \eqref{Int}, the expression for the running of the scalar index given by Eq.~\eqref{Insk} can be written as \cite{German:2020eyq}
\begin{equation}
n_{sk} =\frac{3}{32}r^2 - \frac{1}{2} \delta_{n_s} r -  \frac{1}{4} r \frac{V^{\prime \prime \prime }}{V^{\prime}} \;,
\label{nsk}
\end{equation}
thus, one only has to calculate the last term in Eq.~\eqref{nsk} and eliminate the parameter  $\lambda$ with the result
\begin{equation}
n_{sk} =\frac{3}{64}r^2 - \frac{1}{4} \delta_{n_s} r -  \frac{1}{3} \delta_{n_s}^2 \;.
\label{nskhere}
\end{equation}
Thus, we can also obtain bounds for the running index $n_{sk}$ given by  Eq.~\eqref{nskhere} with the results
\beq
-1/3<\omega_{re}<1/3,\quad 0.0071<r<0.063,\quad \Rightarrow\quad -9.5\times 10^{-4}<n_{sk}<-3.7\times 10^{-4},
\label{casonsk1}
\eeq
and
\beq
1/3<\omega_{re}<1,\quad 0<r<0.0251,\quad \Rightarrow\quad -7.3\times 10^{-4}<n_{sk}<-3.2\times 10^{-4}
\label{casonsk2}.
\eeq
We see that in the first case (Eq.~\eqref{casonsk1}) $n_{sk}$ is of order $10^{-3}$. The scale of inflation $\Delta\equiv V_k^{1/4}$ (given below in $GeV$) can also be bounded as follows
\beq
-1/3<\omega_{re}<1/3,\quad 0.0071<r<0.063,\quad \Rightarrow\quad 9.4\times 10^{15} <\Delta<1.6\times 10^{16},
\label{casodelta1}
\eeq
and
\beq
1/3<\omega_{re}<1,\quad 0<r<0.0251,\quad \Rightarrow\quad 0 <\Delta<1.3\times 10^{16}
\label{casodelta2}.
\eeq
In the Table~\ref{bounds} we collect all these results in a more orderly way.
%%%%%%%%%%%%%%%%%%%%%%%%%%%%%%%%%%%%%%%%%%%%
%%%%%%%%%%%%%%%%%%%%%%%%%%%%%%%%%%%%%%%%%%%%
\section {\bf Conclusions}\label{CON}
We have analytically analyzed the quartic hilltop inflation model in more detail following the suggestion of \cite{Dimopoulos:2020kol}. To achieve this we have implemented a procedure by which the parameters present in the potential that defines the model are eliminated in favor of the observables $n_s$ and $r$. This has allowed us to obtain in a straightforward and simple way the two main equations of \cite{Dimopoulos:2020kol} and we have extended the general discussion in a precise analytical treatment. This procedure has also allowed us to impose bounds on quantities of interest during inflation using a constraint from the reheating period. In this way we have found limits for the parameter $\lambda$ appearing in the potential as well as the running index, the reheating temperature and the inflationary scale. All of these quantities are expressed in terms of $n_s$ and $r$ and their bounds will continue to narrow further as the observations of $n_s$ and $r$ become more precise.

%%%%%%%%%%%%%%%%%%%%%%%%%%%%%%%%%%%%%%%%%%%%
%%%%%%%%%%%%%%%%%%%%%%%%%%%%%%%%%%%%%%%%%%%%
\acknowledgments

We acknowledge financial support from UNAM-PAPIIT,  IN104119, {\it Estudios en gravitaci\'on y cosmolog\'ia}. We also thank the anonymous referee for useful advice.


\begin{thebibliography}{10}


\bibitem{Linde:1984ir}
Andrei~D. Linde.
\newblock {The Inflationary Universe}.
\newblock {\em Rept. Prog. Phys.}, 47:925--986, 1984.

\bibitem{Lyth:1998xn}
David~H. Lyth and Antonio Riotto.
\newblock {Particle physics models of inflation and the cosmological density perturbation}.
\newblock {\em Phys. Rept.}, 314:1--146, 1999.

\bibitem{Baumann:2009ds}
D.~Baumann.
\newblock {Inflation}.
\newblock {\em arXiv:} 0907.5424 [hep-th].

\bibitem{Martin:2018ycu}
Jerome Martin.
\newblock {The Theory of Inflation}.
\newblock In {\em {200th Course of Enrico Fermi School of Physics: Gravitational Waves and Cosmology (GW-COSM) Varenna (Lake Como), Lecco, Italy, July 3-12, 2017}}, 2018.

\bibitem{Linde:1981mu}
Andrei~D. Linde.
\newblock {A New Inflationary Universe Scenario: A Possible Solution of the Horizon, Flatness, Homogeneity, Isotropy and Primordial Monopole Problems}.
\newblock {\em Phys. Lett. B }, 108 (1982)389.

\bibitem{Kinney:1995cc}
Kinney, William H. and Mahanthappa, K.T.
\newblock {Inflation at low scales: General analysis and a detailed model}
\newblock {\em Phys. Rev.}, D53:5455-5467, 1996.

\bibitem{Boubekeur:2005zm}
Boubekeur, Lotfi and Lyth, David. H.
\newblock {Hilltop inflation}.
 \newblock {\em JCAP}, {\bf 07}(2005)010.
 
 \bibitem{Martin:2013tda} 
J.~Martin, C.~Ringeval and V.~Vennin.
\newblock {Encyclopedia Inflationaris}.
\newblock In {\em  Phys.\ Dark Univ.} {\bf 5-6}, 75 (2014).

\bibitem{Akrami:2018odb} 
Y.~Akrami {\it et al.} [Planck Collaboration],
\newblock {Planck 2018 results. X. Constraints on inflation}.
\newblock {\em  arXiv:}\,1807.06211, [astro-ph.CO].

\bibitem{Kallosh:2019jnl}
Kallosh, Renata and Linde, Andrei.
\newblock {On hilltop and brane inflation after Planck}.
 \newblock {\em JCAP}, {\bf 09}(2019)030.

\bibitem{Dimopoulos:2020kol}
Dimopoulos, Konstantinos.
\newblock {An analytic treatment of quartic hilltop inflation}.
\newblock {\em Phys. Lett. B}, 809:135688, 2020.

\bibitem{Lin:2019fdk}
Lin, Chia-Min.
\newblock {Topological Eternal Hilltop Inflation and the Swampland Criteria}.
 \newblock {\em JCAP}, {\bf 06}(2020)015.

\bibitem{Bassett:2005xm} 
B.~A.~Bassett, S.~Tsujikawa and D.~Wands,
\newblock {Inflation dynamics and reheating}.
\newblock {\em Rev.\ Mod.\ Phys.},  {\bf 78}, 537 (2006)

\bibitem{Allahverdi:2010xz}
Rouzbeh Allahverdi, Robert Brandenberger, Francis-Yan Cyr-Racine, and Anupam Mazumdar.
\newblock {Reheating in Inflationary Cosmology: Theory and Applications}.
\newblock {\em Ann. Rev. Nucl. Part. Sci.}, 60:27--51, 2010.
 
 \bibitem{Amin:2014eta}
Mustafa~A. Amin, Mark~P. Hertzberg, David~I. Kaiser, and Johanna Karouby.
\newblock {Nonperturbative Dynamics Of Reheating After Inflation: A Review}.
\newblock {\em Int. J. Mod. Phys.}, D24:1530003, 2014.

 \bibitem{Turner:1983he} 
Turner, Michael S,
\newblock {Coherent Scalar Field Oscillations in an Expanding Universe}.
\newblock {\em Phys. Rev. D}, 28, 1243, 1983.

\bibitem{Liddle:2003as}
Andrew~R Liddle and Samuel~M Leach.
\newblock {How long before the end of inflation were observable perturbations
  produced?}
\newblock {\em Phys. Rev.}, D68:103503, 2003.

 \bibitem{Dodelson:2003vq}
Dodelson, Scott and Hui, Lam.
\newblock {A Horizon ratio bound for inflationary fluctuations}
\newblock {\em Phys. Rev. Lett.}, 91, 131301, 2003.

\bibitem{Liddle:1994dx}
Andrew R. Liddle,  Paul Parsons,  and John D. Barrow, 
\newblock {Formalizing the slow roll approximation in inflation}.
\newblock {\em Phys. Rev. D}, 50: 7222--7232, 1994.

\bibitem{Dai:2014jja}
Liang Dai, Marc Kamionkowski, and Junpu Wang.
\newblock {Reheating constraints to inflationary models}.
\newblock {\em Phys. Rev. Lett.}, 113:041302, 2014.

\bibitem{Munoz:2014eqa}
Julian~B. Munoz and Marc Kamionkowski.
\newblock {Equation-of-State Parameter for Reheating}.
\newblock {\em Phys. Rev.}, D91(4):043521, 2015.

\bibitem{German:2020iwg}
G. Germ\'an.
\newblock {Model independent results for the inflationary epoch and the breaking of the degeneracy of models of inflation}.
 \newblock {\em JCAP}, {\bf 11}(2020)006.

\bibitem{Aghanim:2018eyx} 
N.~Aghanim {\it et al.} [Planck Collaboration],
\newblock {Planck 2018 results. VI. Cosmological parameters},
\newblock {\em arXiv:}\,1807.06209, [astro-ph.CO].

\bibitem{German:2020eyq} 
G.~Germ\'an,
\newblock {Constraints from reheating}.
\newblock {\em arXiv:} 2010.09795, [astro-ph.CO].


 
\end{thebibliography}
\end{document}